\documentclass[Afour,sageh,times]{sagej}

\usepackage{moreverb,url}

\usepackage[colorlinks,bookmarksopen,bookmarksnumbered,citecolor=red,urlcolor=red]{hyperref}

\newcommand\BibTeX{{\rmfamily B\kern-.05em \textsc{i\kern-.025em b}\kern-.08em
T\kern-.1667em\lower.7ex\hbox{E}\kern-.125emX}}

\def\volumeyear{2016}

\begin{document}

\runninghead{Karp et~al.}

\title{Large-Scale Direct Numerical Simulations of Turbulence Using GPUs and Modern Fortran}

\author{Martin Karp\affilnum{1}, Daniele Massaro\affilnum{2}, Niclas Jansson\affilnum{3}, Alistair Hart\affilnum{4}, Jacob Wahlgren\affilnum{1}, Philipp Schlatter\affilnum{2}, and Stefano Markidis\affilnum{1}}

\affiliation{\affilnum{1}Department of Computer Science, KTH Royal Institute of Technology, Stockholm, Sweden\\
\affilnum{2}Engineering Mechanics, KTH Royal Institute of Technology, Stockholm, Sweden
\affilnum{3}PDC Centre for High Performance Computing, KTH Royal Institute of Technology, Stockholm, Sweden
\affilnum{4} Hewlett Packard Enterpise (HPE), UK}

\corrauth{Martin Karp, KTH Royal Institute of Technology, School of Electrical Engineering and Computer Science, Lindstedsvägen 5, 100 44 Stockholm, Sweden}

\email{makarp@kth.se}

\begin{abstract}
We present our approach to making direct numerical simulations of turbulence with applications in sustainable shipping. We use modern Fortran and the spectral element method to leverage and scale on supercomputers powered by the Nvidia A100 and the recent AMD Instinct MI250X GPUs, while still providing support for user software developed in Fortran. We demonstrate the efficiency of our approach by performing the world's first direct numerical simulation of the flow around a Flettner rotor at Re=30'000 and its interaction with a turbulent boundary layer. We present one of the first performance comparisons between the AMD Instinct MI250X and Nvidia A100 GPUs for scalable computational fluid dynamics. Our results show that one MI250X offers performance on par with two A100 GPUs and has a similar power efficiency. 
\end{abstract}

\keywords{GPU, Fortran, Spectral Element Method, Computational Fluid Dynamics, Direct Numerical Simulation, High Performance Computing}

\maketitle

\section{Introduction}
High-performance computing resources drive discoveries in fields of crucial importance for society. A societal issue of increasing importance is environmental sustainability and reducing our energy consumption. Our dependence on non-renewable energy has become an increasingly pressing issue due to climate change and increasing global temperatures~(\citealt{pachauri2007ipcc}). 

\begin{figure}[t]
    \centering
    \includegraphics[scale=0.3]{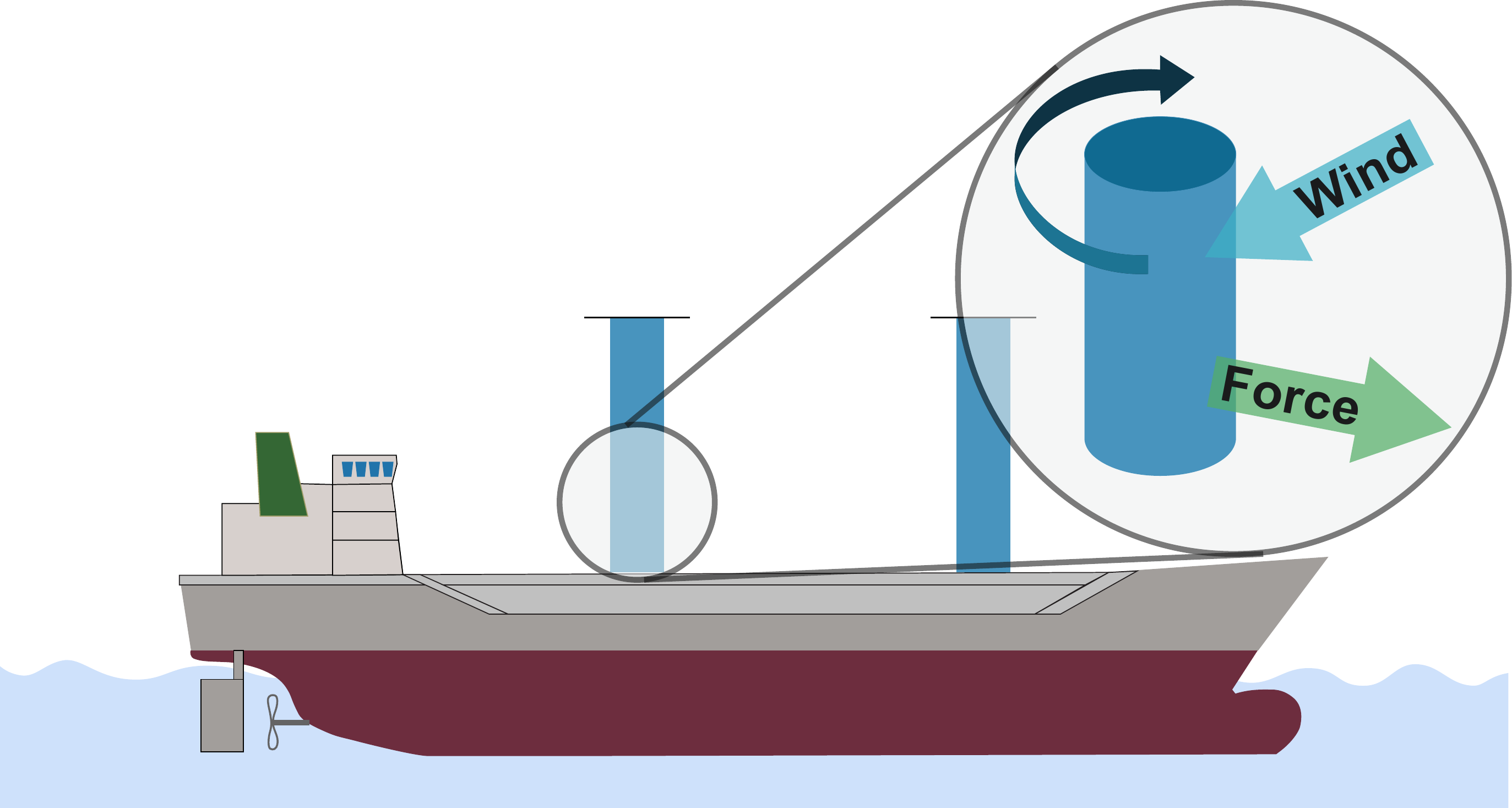}
    \caption{An illustration of a rotor ship which uses Flettner rotors (in blue) to reduce fuel consumption and carbon emissions. We also highlight how the rotor and the wind coming from the side of the ship generates a force directed forward.}
    \label{fig:rotor}
\end{figure}

Global shipping is responsible for more than $2\%$ of all carbon emissions and the largest cargo ships in the world make up a significant portion of these emissions~(\citealt{smith2015co2}). As the number of tonnes of goods transported by waterway is only expected to increase in the coming years, reducing fuel consumption will have a significant impact on both costs, reliablity, and emissions~(\citealt{smith2015co2}). Active research is concerned with several unconventional designs being considered to make ships more energy efficient. One of these approaches is the Flettner rotor, a rotating cylinder that can reduce the fuel consumption in ships by more than 20\% by using the Magnus force, effectively working as a sail~(\citealt{magnus1853,seddiek2021harnessing}). Rotors are now being tested in practice on ships as illustrated in Figure \ref{fig:rotor}. However, as far as numerical simulations go we have so far been limited to less detailed and accurate simulations based on the Reynolds-Averaged Navier-Stokes (RANS) approach~(\citealt{demarco16}). Direct numerical simulations (DNS) that resolve all scales of the flow in both time and space have not been used because of their high computational cost, both with regards to time and power usage. 

Modern computing clusters are now approaching a size and power efficiency where DNS is possible within a reasonable amount of time and number of nodes. Especially, as Graphics Processing Units (GPUs) have become commonplace and outmatch older computing solutions with regards to power efficiency and performance (9 out of the top 10 Green 500 computers are powered by GPUs in November 2021~(\citealt{Green500}). However, while our computer systems are more performant and energy-efficient than ever, computer systems have also grown more complex and heterogeneous.

The increased complexity of modern computer systems has limited the usage of trusted legacy software that scales and performs exemplary on multi-core processors. Therefore, several new frameworks that automate parts of the porting process to heterogeneous hardware, or non-intrusive pragma based approaches, have been introduced~(\citealt{medina2014occa,keryell2015khronos,wienke2012openacc}). However, these packages either require large code changes if the legacy code is written in Fortran~77, or offers relatively low performance. We take another approach, where we use modern Fortran to accommodate various backends, but not be dependent on any single overarching software package. Instead we aim to provide the flexibility to use the most suitable software for a given device. With the Fortran core, we can also efficiently utilize older, verified, CPU code without major changes. We base our solver on the numerical methods and schemes from the spectral element method and Nek5000~(\citealt{nek5000}) and leverage them in Neko -- a portable spectral element framework targeting several computer architectures~(\citealt{jansson2021neko}).

 %to execute the first DNS simulation of a Flettner rotor in a turbulent boundary layer. For our simulation we are able to leverage . As these GPUs power a significant portion of current and future supercomputers, we compare them both with regards to power efficiency and performance for high-order, large-scale computational fluid dynamics. 
Neko can both exploit the computing power of modern accelerators and use older tools for Nek5000. Using the spectral element method, Neko provides high-order convergence on unstructured meshes while having a low communication cost and high performance. In this work, we present our development and usage of Neko to leverage two state-of-the-art GPUs, the Nvidia A100 and the very recent AMD Instinct MI250X. In addition, we interface with older toolboxes written for Fortran~77. We present the efficiency of our solver by making the first DNS of a Flettner rotor in a turbulent boundary layer. We illustrate how our developments lead to energy and time savings, both by strong-scaling and reducing the run time of the simulation, but also how a solver in modern Fortran can make the porting process to new hardware easier.

Our contribution is the following:
\begin{itemize}
    \item We illustrate the software and algorithmic considerations in developing Neko, a  highly scalable solver that caters both to increasingly heterogeneous supercomputers and legacy user software.
    \item We present the first direct numerical simulation of a Flettner rotor and its interaction with a turbulent boundary layer at Reynolds number of 30 000.
    \item We show the scaling, performance, and power measurements of our simulations on both the new AMD Instinct MI250X and Nvidia A100 GPUs, showing a matching power efficiency and  that one MI250X performs similarly to two A100s. 
\end{itemize}

The paper is organized as follows, in Section II.  we present Neko, its numerical methods and software design. In Section III. we briefly cover the physics of a Flettner rotor before we describe in Section IV. our simulation setup and computer platforms used for our experiments. In Section V. we present the simulation results and compare the performance and power efficiency of our computer platforms. In the two final sections we cover the related work and make our concluding remarks. 
%Should we use the real name?
\section{Neko}
%Modern, roots in Nek5000, accomodate different backends
Neko is a spectral element framework targeting incompressible flows. With its roots and main inspiration in Nek5000, a CFD solver widely acclaimed for its numerical methods and scalability on multicore processors~(\citealt{tufo1999terascale}), Neko is a new code making use of object-oriented Fortran to control memory allocation and allow multi-tier abstractions of the solver stack. Neko has been used to accommodate various hardware backends ranging from general-purpose processors, accelerators, vector processors~(\citealt{hpcasia21}), as well as (limited) FPGA support~(\citealt{karp2021High,karp2022HpcAsia}). In our work, we extend and optimize Neko to perform a large-scale DNS of practical industrial relevance on GPUs. The spectral element method used in Neko is a high-order, matrix-free, finite element method which provides high order convergence on unstructured meshes. Here we provide a brief overview of Neko, the spectral element discretization, time-stepping schemes, numerical solvers, and an overview of the overarching software structure.

\subsection{Discretization}
We integrate the incompressible, non-dimensional, Navier-Stokes equation in time
\begin{align}
    \nabla  \cdot  \mathbf{v} &= 0, \label{eq:incom}\\ 
    \frac{\partial \mathbf{v}}{\partial t} + \mathbf{v}\cdot \nabla \mathbf{v} &= - \nabla p + \frac{1}{Re}\nabla^2 \mathbf{v}\label{eq:mom} + \mathbf{F}
\end{align}
where $Re = \frac{UL}{\nu}$ is a Reynolds number defined by a characteristic velocity $U$, length scale $L$, and kinematic viscosity $\nu$, $\mathbf{v}$ is the instantaneous velocity field, $p$ the non-dimensional pressure and $\mathbf{F}$ an external forcing/source term. We solve the problem on a domain $\Omega$ with suitable boundary conditions on $\partial \Omega$. As this system of equations is coupled, we use the so called $P_N-P_N$ splitting scheme to separate the pressure and velocity solves at each time step. This splitting scheme was originally proposed by \cite{karniadakis1991high} and has shown to have the same order of convergence as the time stepping scheme that is used, with the error being primarily isolated to the boundaries of the domain. The $P_N-P_N$ scheme discretizes the momentum equation in time in the following fashion
\begin{equation}\label{eq:splitting}
 \frac{\gamma_0 \mathbf{v}^{n+1}}{\Delta t} = \frac{\mathbf{\hat{v}}}{\Delta t} - \nabla p^{n+1} + \frac{1}{\text{Re}} \nabla^2 \mathbf{v}^{n+1}
\end{equation}
with 
\begin{equation}
    \hat{\mathbf{v}}\frac{}{} = \Delta t\sum^{J_e-1}_{q=0} \beta_q \mathbf{v}^{n-q}\cdot \nabla \mathbf{v}^{n-q}+\sum_{q=0}^{J_i-1}\alpha_q\mathbf{v}^{n-q} + \mathbf{F}.
\end{equation}
We use an explicit scheme of order $J_e$ for the nonlinear terms and an implicit scheme of order $J_i$ for the linear terms. In our case we use a standard backward differentiation scheme of order three for the implicit terms and a extrapolation scheme of order three for the explicit terms. To decouple the velocity and pressure solves we take the divergence of \eqref{eq:splitting} and use incompressiblity, our explicit treatment of $\mathbf{v}$ and the identity
\begin{equation}
    \nabla^2\mathbf{v}^{n+1}=\nabla(\nabla\cdot\mathbf{v}^{n+1})-\nabla\times(\nabla\times\mathbf{v}^{n+1})
\end{equation}
to arrive at a pressure-Poisson equation for the pressure 
\begin{equation}\label{eq:pressure}
    -\nabla^2p^{n+1} = -\frac{\nabla \cdot \mathbf{\hat{v}}}{\Delta t} +\frac{1}{\text{Re}}\nabla \cdot (\nabla \times \omega )
\end{equation} with $\omega$ being the vorticity computed explicitly
\begin{equation}
\omega = \sum^{J_e-1}_{q=0} \beta_q (\nabla \times \mathbf{v}^{n-q}).
\end{equation}
With this approach, we have decoupled the pressure and velocity solve and each time step can thus be computed by first solving the system for the pressure according to \eqref{eq:pressure} followed by using the resulting pressure field $p^{n+1}$ in equation \eqref{eq:splitting} and obtaining the velocity at step $n+1$. { Crucial to this splitting approach is choosing suitable boundary conditions for the pressure. It is found that the temporal error of the splitting scheme is highest along the boundary and directly depends on properly imposed pressure boundary conditions. Using Neumann boundary conditions it is possible to limit the temporal splitting error for the pressure and divergence to $\mathcal{O}(\Delta t^{J_e})$ and thus limit the splitting error for the velocity to $\mathcal{O}(\Delta t^{J_e+1})$ as shown in \cite{orszag1986boundary,karniadakis1991high}. In addition, it can be shown that the error in the divergence of the velocity at step $n+1$ decays exponentially from the boundary proportionally to $e^{-s/l}$, where $s$ is a coordinate normal to the boundary and $l=\sqrt{\gamma_0 \nu \Delta t}$ is the so-called numerical boundary layer thickness. As $l\rightarrow 0$ for $Re\rightarrow\infty$ the error decays rapidly from the boundary for highly turbulent flows~(\citealt{karniadakis1991high}}). Unique for this splitting scheme, compared to the more conventional $P_N-P_{N-2}$ splitting for the spectral element method presented by \cite{maday1989spectral} is that the velocity and pressure solve can be done with the same function space, and thus use the same polynomial degree $N$. With our temporal discretization in place, we apply our spectral element discretization in space.  

The spectral element method is used because of its high-order convergence and high accuracy with relatively few grid points in space compared to e.g. finite volume methods, while still accommodating unstructured meshes~(\citealt{deville2002high}). The linear system is evaluated in a matrix-free fashion, using optimized tensor-operations, which yields a high operational intensity, performance, and low communication costs. These aspects make the spectral element method well suited for DNS in complex geometries.

For the spectral element discretization we consider the weak form of the Navier-Stokes equations and decompose the domain into $E$ non-overlapping hexahedral elements $\Omega = \bigcup^E_e\Omega^e$. On these elements we represent the solution with polynomial basis functions based on the Legendre polynomials of degree $N$ interpolated on the Gauss-Lobatto-Legendre (GLL) points $\xi_i, i = 0, N$. In 3D the number of GLL points per element is $(N+1)^3$. A function on the reference element can be represented according to
\begin{equation}
    u^e(\xi, \eta, \gamma) = \sum_{i,j,k=0}^N u^e_{i,j,k} l_i(\xi)l_j(\eta)l_k(\gamma)
\end{equation}
where $l_i$ are the polynomial basis functions and $\xi, \eta, \gamma$ are the location within the reference element. As we use non-overlapping hexahedral elements, differentiation can be evaluated efficiently through tensor-products~\cite{orszag1979spectral}. For instance, the weak form of the Laplace operator can written as
\begin{equation}
\label{eq:disclap}
    (\nabla v, \nabla u) = \sum^E_e (v^e)^T \mathbf{D}^T \mathbf{G}^e\mathbf{D} u^e = \sum^E_e (v^e)^T A^e u^e
\end{equation}
where we introduce the differentiation matrices $\mathbf{D}$ and the geometric factors $\mathbf{G}^e$. These two matrices are defined as
\begin{equation*}
\mathbf{D}_{ii} = \frac{\text{d} l_i(\xi)}{\text{d}\xi}\bigg\rvert_{\xi_i}, \quad (\mathbf{G}^e_{st})_{ijk}=\rho_i\rho_j\rho_k\mathcal{J}_{ijk} \sum_{l=1}^3 \frac{\partial r_s}{\partial x_l}\frac{\partial r_t}{\partial x_l}
\end{equation*}
where $\mathcal{J} = \frac{\partial \mathbf{x}}{\partial r}$ is the Jacobian for the mapping to the reference element. We have that $\mathbf{G}^e_{st}=\mathbf{G}^e_{ts}$ and thus we have that $\mathbf{G}^e$ is symmetric and contains 6 non-trivial values per Gauss-Lobatto-Legendre (GLL) point. In Neko, the global system $A^e$ is never formed as this would be prohibitely expensive, but $A^e$ is instead evaluated in a matrix-free fashion (\citealt{deville2002high}). For this we introduce the  local representation $u_L = \mathcal{Q}u$ where $\mathcal{Q}$ is known as the scatter operator. The local representation $u_L$ contains duplicate values along element boundaries, but enables for efficient tensor-operations. By ensuring that 
\begin{equation}
    u^e_{ijk} = u^{e^\prime}_{i^\prime j^\prime k^\prime} \quad \text{if} \quad \mathbf{x}^e_{ijk} = \mathbf{x}^{e^\prime}_{i^\prime j^\prime k^\prime}
\end{equation}
and using the local representation we can therefore evaluate the Laplace operator as
\begin{equation}
    \sum^E_e (v^e)^T A^e u^e = (\mathcal{Q}v)^TA_L\mathcal{Q}u
\end{equation}
with $A_L$ being the block diagonal matrix of $A^e$. The final discrete system that we solve for can thus be written as
\begin{equation}
    \mathcal{QQ}^T A_L u_L = Au_L.
\end{equation}
 A similar approach is applied to the Helmholtz equation for the velocity arising from the implicit treatment of the viscous term. Using this discretization and time splitting we have all the components to set up our flow case. To solve the systems for the velocity and pressure we use heavily optimized numerical solvers. 
%Weak formulation, PN-PN, Timestepping scheme, I want this to be rigorous. EXTEND THIS SECTION

\subsection{Numerical Solvers}
The pressure solve is the main source of stiffness in incompressible flow and because of this, we use a restarted generalized minimal residual method (GMRES)~(\citealt{saad1986gmres}) combined with a preconditioner based on an overlapping additive Schwarz method and a coarse grid solve~(\citealt{lottes2005hybrid}). In Neko, the coarse grid solve is done with 10 iterations of the preconditioned conjugate gradient (P-CG) method. This type of preconditioner significantly reduces the number of iterations necessary for convergence at the expense of increased work per iteration.  

For the velocity solver, we utilize P-CG with a block-Jacobi preconditioner. For both the velocity and pressure solve we utilize a projection scheme where we project the solution of the current time step on previous solution vectors to decrease the start residual significantly. This has been observed to reduce the number of iterations for the Krylov solver significantly (\citealt{fischer1998projection}). All of our solvers are designed with locality across the memory hierarchy in mind, to be able to as efficiently as possible use modern GPUs with a significant machine imbalance. Parts of this optimization process and the theoretical background is described in \cite{karp2022reducing}.

\begin{figure*}
    \centering
    \includegraphics[trim=0.15cm 0.8cm 8.2cm 2.5cm,clip,width=1.0\linewidth]{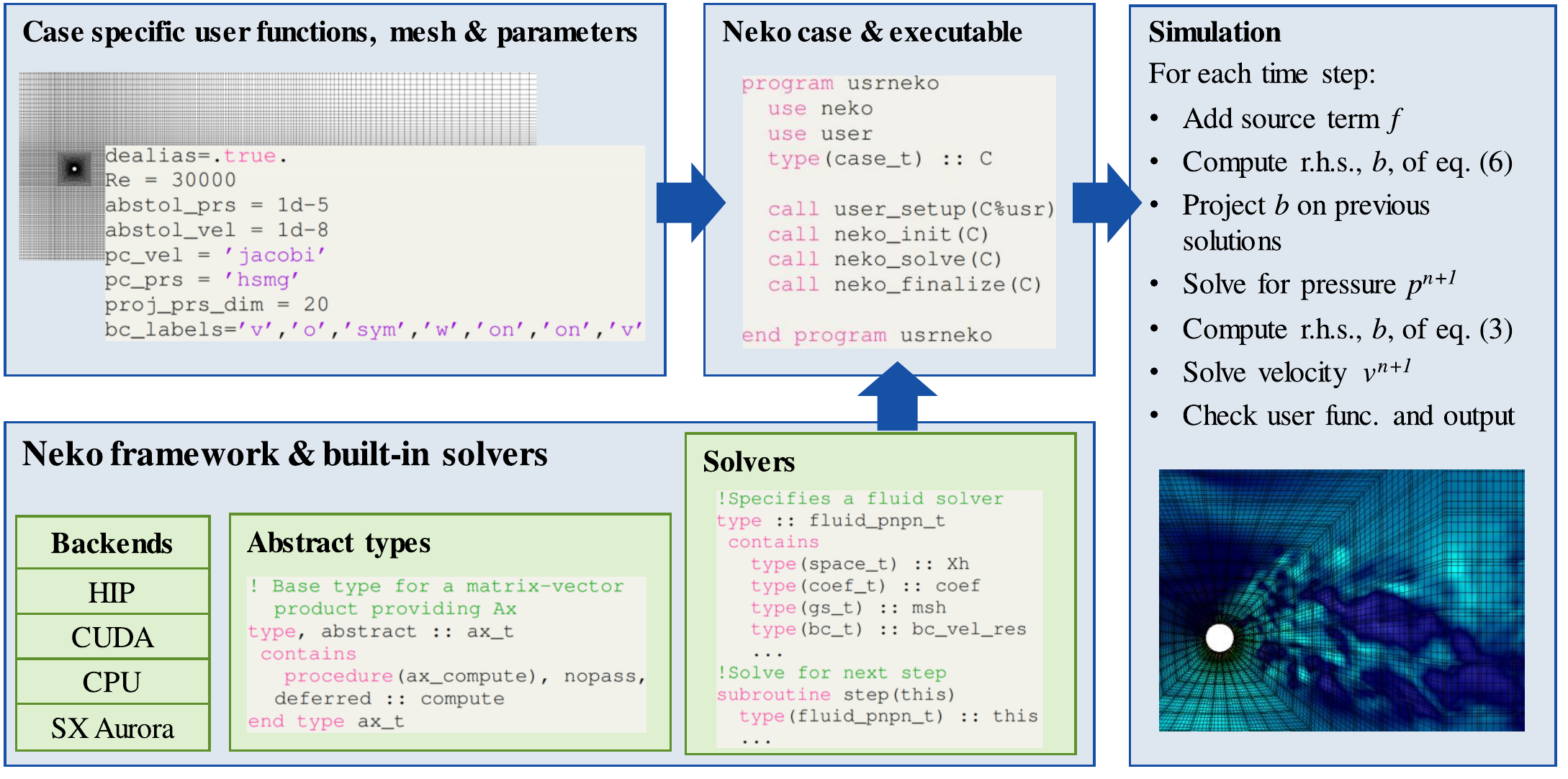}
    \caption{A chart illustrating the structure of the Neko framework. The Neko framework, with various abstract types and backends make up different solvers which are used to solve a case. The case is defined by user provided functions, as well as the input parameters and mesh. In our case, the simulation is performed using the $P_N-P_N$ method to perform a DNS of a Flettner rotor.}
    \label{fig:neko_chart}
\end{figure*}

We illustrate how each time step is computed on a high level in Figure \ref{fig:neko_chart}. Here we also illustrate the software structure of Neko, which we will cover more in-depth in the next section.

\subsection{Software Design}
%Krylov solver $\rightarrow$ pipelined CG and pipelined GMRES?
%Convergence $\rightarrow$ Relative?
%Precon $\rightarrow$ CA-CG crs grid solver

%Device layer, flexiblity, lots of control for developers, high lever for users

The primary consideration in Neko is how to efficiently utilize different computer backends, without re-implementing the whole framework for each backend and while maintaining the core of the solver in modern Fortran. We solve this problem by considering the weak form of the equations used in the spectral element method. The weak formulation allows us to formulate equations as abstract problems which enable us to keep the abstractions at the top level of the software stack and reduce the amount of platform-dependent kernels to a minimum. The multi-tier abstractions in Neko are realized using abstract Fortran types, with deferred implementations of required procedures. For example, to allow for different formulations of a simulation's governing equations, Neko provides an abstract type \verb|ax_t|, defining the abstract problem's matrix-vector product. The abstract type is shown in Figure \ref{fig:neko_chart} under Abstract types and requires any derived, concrete type to provide an implementation of the deferred procedure \verb|compute|. This procedure would return the action of multiplying the stiffness matrix of a given equation with a vector. For the Poisson equation, the \verb|compute| subroutine corresponds to computing expression \eqref{eq:disclap}.

%\begin{figure*}
%    \centering
%    \includegraphics[trim=0.1cm 0.22cm 8.5cm 4.0cm,clip,width=1.0\linewidth]{picture_neko-5.pdf}
%    \caption{A flowchart illustrating the structure of the Neko framework. The Neko library, with various abstract types and backends make up different solvers which are used to solve a case. The case is defined by user provided functions, as well as the input parameters and mesh. In our case, the simulation is performed using the $P_N-P_N$ method to perform a DNS of a Flettner rotor.}
%    \label{fig:neko_chart}
%\end{figure*}

%\begin{figure}
%    \centering
%    \includegraphics[trim=0.1cm 1cm 13.3cm 0.2cm,clip,width=1.0\linewidth]{picture_neko_revision.pdf}
%    \caption{A flowchart illustrating the structure of the Neko framework. The Neko library, with various abstract types and backends make up different solvers which are used to solve a case. The case is defined by user provided functions, as well as the input parameters and mesh. In our case, the simulation is performed using the $P_N-P_N$ method to perform a DNS of a Flettner rotor.}
%    \label{fig:neko_chart}
%\end{figure}
In a typical object-oriented fashion, whenever a routine needs a matrix-vector product, it is always expressed as a call to \verb|compute| on the abstract base type and never on the actual concrete implementation. Abstract types are all defined at the top level in the solver stack during initialization and represent large, compute-intensive kernels, thus reducing overhead costs associated with the abstraction layer. 

Furthermore, this abstraction also accommodates the possibility of providing tuned matrix-vector products (\verb|ax_t|) for specific hardware, only providing a particular implementation of \verb|compute| without having to modify the entire solver stack. The ease of supporting various hardware backends is the key feature behind the performance portability of Neko. 

However, a portable matrix-vector multiplication backend is not enough to ensure performance portability. Therefore, abstract types are used to describe a flow solver's common parts. These abstract types describe all the parts necessary to compute a time step and provides deferred procedures for essential subroutines. These building blocks then make up the entire solver used to calculate one time step. We illustrate this structure in Figure \ref{fig:neko_chart} where a solver such as \verb|fluid_pnpn_t| implements the deferred subroutine \verb|step|. The solver is then ready to solve a particular case. Each case (\verb|case_t|) is defined based on a mesh (\verb|mesh_t|), parameters (\verb|param_t|), together with the abstract solver (\verb|solver_t|),  later defined as an actual extended derived type, such as \verb|fluid_pnpn_t|, based on the simulation parameters. The abstract solvers contains a set of defined derived types necessary for a spectral element simulation, such as a function space (\verb|space_t|), coefficients (\verb|coef_t|), and various fields (\verb|field_t|).  Additionally, they contains further abstract types for defining matrix-vector products (\verb|ax_t|) and gather-scatter $QQ^T$ kernels (\verb|gs_t|). Each of these abstract types is associated with an actual implementation in an extended derived type, allowing for hardware or programming model specific implementations, all interchangeable at runtime. This way, Neko can accommodate different backends with both native and offloading type execution models without unnecessary code duplication in the solver stack.

\subsection{GPU Implementation Considerations}

{ As Neko provides abstractions on several levels it is possible to utilize different libraries to leverage accelerators.} In our work, we do not rely on vendor-specific solutions (e.g. CUDA Fortran) or directives-based approaches; instead, we exploit the device abstraction layer to manage device memory, data transfer, and kernel launches from Fortran. Behind this interface, Neko calls the native accelerator implementation written in, e.g. CUDA, HIP or OpenCL. In our work we extend and optimize the CUDA and HIP backends in Neko.

{ As we extend several features of the device layer in Neko in this work, a few aspects of the code development are important to point out regarding the GPU backend. In particular, we observed a considerable difference in how HIP and CUDA kernels map to their respective platform. As several important kernels in Neko follow a similar tensor-product structure to the kernels considered by \cite{swirydowicz2019acceleration}, and perform close to the memory-bound roofline on Nvidia GPUs, the same kernels ported to HIP needed to be changed to better exploit AMD GPUs. We found that a 1D thread structure and manual calculation of indices in a thread block enabled higher performance on the AMD Instinct MI250X, rather than launching a 2D/3D thread block as we did previously on Nvidia. It would also appear that there is a difference in the order the threads execute their operations on AMD, meaning that we at one point needed to synchronize the thread-block in HIP, while the same code performed deterministically on Nvidia GPUs. These differences between CUDA/Nvidia and HIP/AMD warrant further and a more systematic investigation as it is not immediately evident why we observe such differences. 

One contribution of the GPU backend in Neko is that we limit data movement to and from the device as far as possible. In practice, this means that only data needed for communication over the network or to do I/O is sent back to the host. As the bandwidth between host and device is significantly lower than that of the high-bandwidth memory (HBM) on the GPUs, limiting data movement to the host is essential to obtain high performance. However, this also means that the HBM memory of the GPUs must fit all the data necessary to carry out the simulation. While it might be beneficial to utilize both CPU and GPUs in the future, the GPU nodes evaluated in this work have significantly more powerful GPUs than the host CPU with regards to both flop/s and memory bandwidth (more than $10\times$). We found that the most effective way to leverage this computing power was to use the host only as a coordinator and offload the entire calculation to the GPUs.

In addition to limiting the data movement to the host, we also avoid data movement from the GPU global memory as far as possible. We do this by merging kernels and by utilizing shared memory and registers in compute heavy kernels. For modern GPUs, the spectral element method is in the memory-bound domain as discussed by \cite{kolev2021efficient} and optimizing the code for temporal and spatial locality is our main priority when designing kernels for the GPU backend in Neko.}

Using the device abstraction layer, from the user perspective, it is possible to write user-specified functions without having to implement handwritten accelerator code e.g. CUDA or HIP. This enables tools from Nek5000 to be used { on the device} with relatively minor changes as we introduce \verb|device| kernels with the same functionality as the original subroutines in Nek5000. Neko provides support for different backends for the gather-scatter operation and $Ax$ as previously mentioned, but also for mathematical operators such as the gradient and curl. With the device layer, we can with only minor changes incorporate verified toolboxes and functions necessary for our flow case into Neko and run them { directly on the GPU without sending data back to the host}. 

%{ Kernels AMD vs Nvidia, data movement, heterogeniety}

\subsection{Modern Fortran}
As is clear from the description of Neko, modern Fortran { (e.g. Fortran 03, 08, 18)} offers similar object-oriented features to other programming languages such as C++~(\citealt{reid2008new}). Most importantly, modern Fortran supports dynamic memory allocation. Other features of Fortran are the native support for the quad data-type, FP128, that all arrays are by default not aliased, pure functions, and that modern Fortran both supports older Fortran code and can interface with C-code. As Fortran is tailored to high-performance numerical computations it caters to the functionalities necessary to perform high-fidelity flow simulations.

\begin{figure*}[t]
    \centering
    \includegraphics[trim=1cm 3cm 1.3cm 1.3cm,clip,scale=0.55]{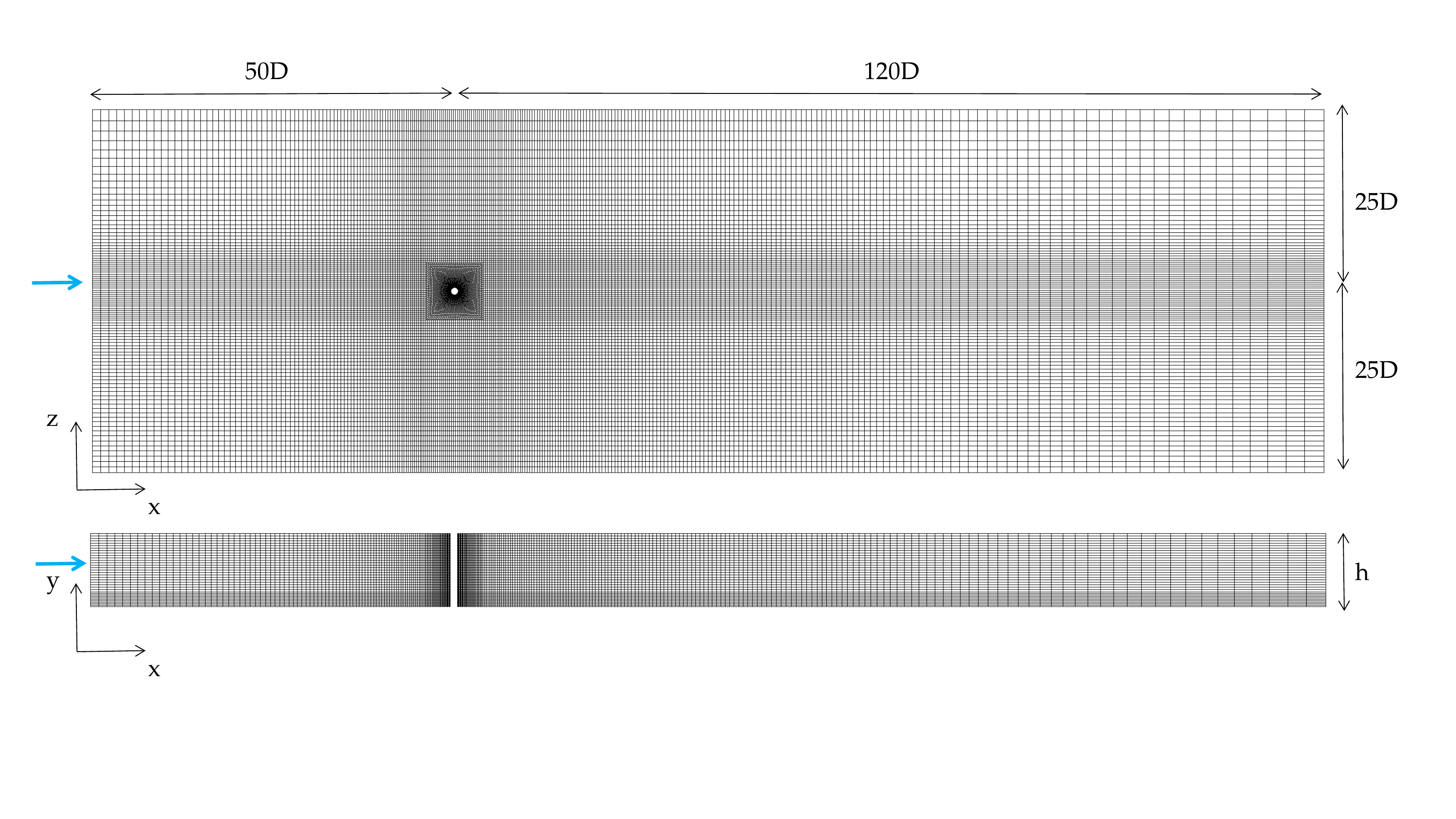}
    \caption{The spectral elements grid is shown. The domain sizes are reported in terms of cylinder diameter $D$ and open channel height $h$ ($\gamma=h/D$=10). The blue arrows indicate the flow direction.}
    \label{fig:mesh}
\end{figure*}
\section{Flettner Rotor}
A Flettner rotor, named after the German inventor Anton Flettner, uses the Magnus effect to generate lift, a force perpendicular to the incoming flow.  The physical phenomenon has been known in the literature since the nineteenth century and was introduced first by \cite{magnus1853}. The Magnus effect consists of the generation of sidewise force by a rotating object, \emph{e.g.} a cylinder or sphere, submerged in a flow. The force results from the velocity changes induced by the spinning motion. The rotation leads to a pressure difference at the cylinder's opposite sides, generating a force pointing from the high to the low pressure side.  In a rotor ship, the cylinder stands vertically and is mechanically driven to develop lift in the direction of the ship, as illustrated in Figure \ref{fig:rotor}. Over the last century, various designs using this effect have been proposed for both ships and airplanes, but with limited success~(\citealt{seifert2012review}). Recently, Flettner rotors have gained renewed interest as a potential alternative to sails to reduce the fuel consumption of ships on our way towards net-zero carbon emissions. This is already being tested in practice for a select number of shipping routes. Because of Flettner rotors' promise in reducing carbon emissions, DNS of a Flettner rotor and its interaction with a turbulent boundary layer are of fundamental interest.

\section{Experimental Setup}
We describe here our precise problem description to accurately capture the physics of a Flettner rotor in a turbulent boundary layer. In addition, we detail the computational setup for our scaling runs on the GPUs. We also describe our baseline CPU platform and provide some further information regarding our final production run that was used to collect the simulation results.

\subsection{Flow Configuration}
In marine applications, the Flettner rotor is usually placed on the deck of the ship and hit by an incoming flow, which is reasonably modeled by a  turbulent boundary layer with a certain thickness, typically on the order of the rotor height. In the current model, we simulate such a configuration as an open channel flow where the Flettner rotor is located at the origin of our domain. The reference system is oriented with the $y$-axis along with the cylinder (vertical) and the $x$ and $z$ being the streamwise and spanwise directions respectively. The flow is fully characterized by three non-dimensional parameters:
\begin{itemize}
    \item the Reynolds number based on the center-line velocity $u_{cl}$ and the height $h$ of the open channel: $Re_{cl}=u_{cl} h/\nu$, where $\nu$ is the kinematic fluid viscosity
    \item the ratio between the height $h$ and the cylinder diameter $\gamma=h/D$
    \item the ratio between the center-line and the spinning cylinder velocity $\alpha=u_{sp}/u_{cl}$.
\end{itemize}
We set those parameters as $Re_{cl}=30~000$, $\gamma=10$, and $\alpha=3$, aiming to reproduce a realistic configuration, within the limits of DNS where all the scales are simulated and no turbulence model is introduced. 

\begin{figure}
    \centering
    \includegraphics[trim=9cm 3cm 10cm 3cm,clip,width=0.42\textwidth]{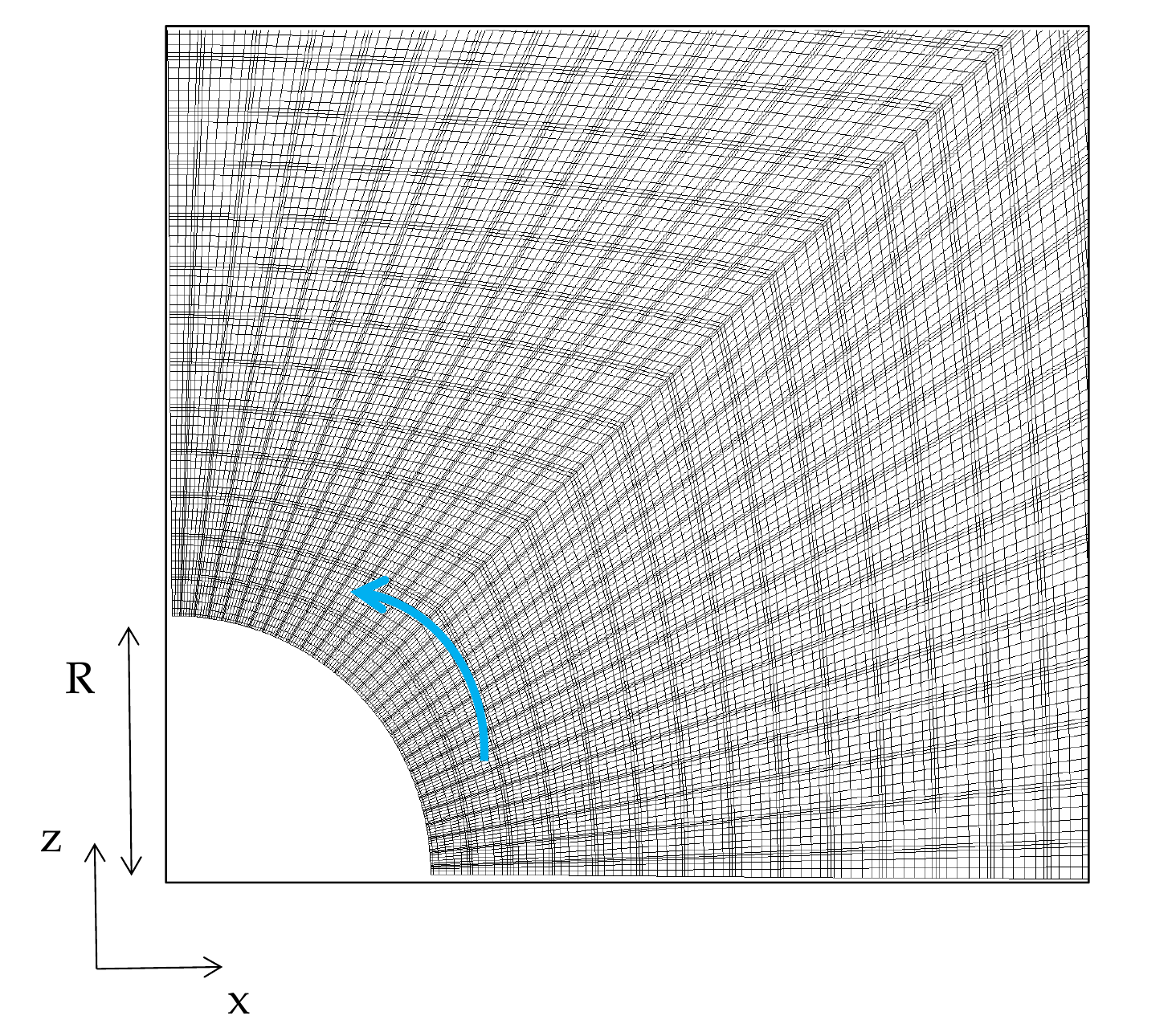}
    \caption{The figure shows the Gauss-Lobatto-Legendre points around the cylinder. The cylinder rotates along the vertical $y$ axis, and the blue arrow indicates the direction of rotation.}
    \label{fig:gll_points}
\end{figure}

The mesh is Cartesian with a block-structured configuration, taking into account the resolution requirements in the different regions and the flow physics. It counts $930070$ spectral elements, which turns into $n\approx 0.48$ billion unique grid points since the polynomial order is $N=7$  { and we have $8^3$ GLL points per element. We show the spectral element mesh with domain dimensions, without GLL points, in Figure \ref{fig:mesh}.} { Close to the cylinder, high resolution is necessary to properly resolve the smallest scales in the boundary layer and we show a snapshot of the large number of GLL points in Figure~\ref{fig:gll_points}.} For the simulation, a time step of $\Delta t^* = 2\cdot10^{-5}$ is used, corresponding to a Courant–Friedrichs–Lewy (CFL) number around $0.41$. { With regards to our iterative solvers, we use a residual tolerance of $10^{-5}$ for the pressure and $10^{-8}$ for the velocity. The choice of tolerances is case-dependent. We consider different orders of magnitude as the pressure solver is the most time-consuming. However, the relaxed tolerance for the pressure is still small enough compared to other error sources.}  The grid is designed to avoid any blockage effects on the rotor, \emph{i.e.} the domain sizes are adequately, and conservatively, large: $(-50D,120D)$ in $x$, $(-25D,25D)$ in $z$ and $(0,10D)$ in $y$, where $x,z,y$ are the streamwise, spanwise and vertical directions  respectively. 

At the inflow ($x=-50D$) the Dirichlet boundary condition (BC) prescribes a turbulent channel velocity profile with the power-law $u_x/u_{cl}=(y/h)^{1/7}$. Unlike the parabolic laminar profile, it is flatter in the central part of the channel and drops rapidly at the walls. On the rotor, a Dirichlet-like BC is used as well, prescribing the wall impermeability and setting a rotational velocity $u_{\theta}= u_{sp}=\alpha  u_{cl}$ with $\alpha=3$ and $u_{\rho}=0$. Here, the BC is expressed in cylindrical coordinates, being $u_{\theta}$ and $u_{\rho}$ the velocities in the azimuthal and radial direction, respectively. To avoid incompatibility conditions at the base of the cylinder the spinning velocity is smoothed as a function of $y$
\begin{equation}
\begin{split}
s&(y=0)=0, \\
s&(0<y<\delta)=u_{sp}/(1+e^{1/(q-1)+1/q})\\
s&(y\geq \delta)=u_{sp}
\end{split}
\end{equation}    
where $q=y/\delta$ and $\delta=0.02h$. The outflow  at $x=120D$ consists of natural boundary condition $ (-p \mathbf{I} + \nu \nabla \mathbf{u})\cdot \mathbf{n} = 0 $. We consider an infinitely long cylinder in the $y$ direction, as a consequence symmetric Robin (or mixed) boundary conditions are prescribed at the top surface ($\mathbf{u} \cdot \mathbf{n} = 0$ with $(\nabla \mathbf{u} \cdot \mathbf{t}) \cdot \mathbf{n}= 0$). For the spanwise boundaries mixed conditions allowing transportation are used, similar to the open boundary, but prescribing no velocity changes in non-normal directions.

To keep the front streamwise extent of the computational domain as short as possible, reduce the computational cost, and sustain the incoming turbulence, the laminar-turbulent transition is initiated via the boundary layer tripping introduced by \cite{schlatter12}. It consists of a stochastic forcing term which is added in a small elliptical region close to the wall. The region is centered along a user-defined line at $x_0=-4.5$ and it extends from $z=-1$ to $z=1$. The implementation follows \cite{schlatter12}, using a weak random volume force that acts in the $y$ direction. The term which enters the Navier-Stokes equation is:
\begin{equation}
    F_2 = g(z,t)\cdot \exp\left(\frac{(x-x_0)^2}{l_x^2} - \frac{y^2}{l_y^2}\right)
\end{equation}
where $l_x$ and $l_y$ are the spatial Gaussian attenuation of the forcing region. The function $g(z,t)$ consists of two terms, corresponding to steady and unsteady perturbations, with amplitudes $T_s$ and $T_u$ respectively. The form is the following:
\begin{equation}
\begin{split}
\label{eq:tripping}
g(z, t)&= T_s g(z)\\
&+ T_u\left[(1-b(t)) h^{i}(z)+b(t) h^{i+1}(z)\right]
\end{split}
\end{equation}
with $i =\text{int}(t / t_s)$, $p =t /t_s - i$, and $b(t) =3 p^{2}-2 p^{3}$. Third-order Lagrangian interpolants with time scale $t_s$ are used for the temporal fluctuations and the forcing has a continuous first order derivative in time. The functions $g(z)$ and $h(z)$ are Fourier series of unit amplitude with $N_m$ random coefficients. The noise is generated with a uniform distribution over all frequencies lower than the cutoff frequency, corresponding to $2\pi/t_s$. The values of the parameters has to be properly chosen, here we use $T_s=0$, $T_u=0.3$, $t_s=0.14$ and $N_m=40$.

This tripping procedure has been used extensively in Nek5000 (see \cite{tanarroetal19,atzorietal21}) and keeps the same original implementation of the pseudo-spectral code SIMSON (\citealt{Chevalier07}). We implement the interpolation on the GPU using the Neko device layer to avoid unnecessary data movement to and from the device. This can be done in Neko without making too many changes to the original implementation and we can rely on a tripping procedure that is well-used and validated in our simulation.

\subsection{Computational Setup}
%alvis, amd setup
We perform our experiments on two different GPU clusters, Alvis at C3SE at Chalmers University of Technology in Gothenburg (Sweden) and on an internal HPE Cray EX system. We present the technical details of both computer setups in Table~\ref{tbl:exp_setup}. The largest difference between the two setups is the Peak double precision (FP64) performance of their respective GPUs, but only a factor of two difference for the available memory bandwidth. { As each MI250X consists of two graphics compute dies (GCDs), each GCD  corresponding to one logical GPU, we will throughout the discussion compare logical GPUs, comparing one GCD to one A100.} We note that the HPE Cray EX system is an internal engineering test system at HPE. Although our results are in line with expectations, it is possible that the performance or scalability are affected by temporary hardware settings and configuration.

To provide a baseline, we also make CPU measurements on Dardel, also a Cray HPE EX system at PDC at KTH Royal Institute of Technology. Each node on Dardel is equipped with 2 AMD EPYC 7742 CPUs with 256 GB DDR4 memory. For compilation, we used the Cray Compiler Environment (CCE) 11.0.4 and Cray MPICH 8.1.11. The interconnect on Dardel is Slingshot 10.

\begin{table*}[]
    \centering
    \begin{tabular}{lll}\toprule
       Setup  & Alvis & HPE Cray EX  \\\midrule
       GPU  & 4 Nvidia A100 SXM4 & 4 AMD Instinct MI250X\\
       CPU & 2 Intel Xeon Gold 6338& AMD EPYC 7A53\\
       DRAM & 512GB DDR4 & 512GB DDR4\\
       MPI & OpenMPI 4.0.5 & Cray MPICH 8.1.14 \\
       Interconnect & Mellanox ConnectX-6  & HPE Slingshot Interconnect  \\
        & (2x200 Gb/s) &200 GbE NICs(4x200 Gb/s)\\
       Compiler & GCC 10.2& CCE 13.0 \\
       GPU Driver & 510.39.01 & 5.11.32 \\
       CUDA/ROCM & CUDA 11.1.1 & ROCM 4.5.2 \\
       Bandwidth & 1550 GB/s & 3277 GB/s\\
       GPU Peak FP64 & 9.746 TFLOPS & 47.87 TFLOPS\\
       GPU TDP & 400W & 560W  \\
       \bottomrule \\
    \end{tabular}
    \caption{Hardware and software details for our GPU platforms. Hardware listed is per node. Bandwidth and performance is per GPU. FP64 from tensor cores omitted.}
    \label{tbl:exp_setup}
\end{table*}
\subsection{Experimental Methodology}
For our scaling runs we scale between 8 and 32 Alvis nodes each with four A100 GPUs. We measure the performance over 500 time steps and use the last 100 to collect the average wall time per time step as well as the standard deviation. The same procedure is used on the HPE cluster where we scale from $4-16$ nodes, each equipped with four MI250X GPUs. Since they are dual chip GPUs, we use eight MPI ranks per node. On Dardel we scale from $16-64$ nodes, each with two CPUs, using one MPI rank per core. As the flow case is large, we could not measure the performance at a fewer number of nodes.

We also compare the power consumption of the two different GPU clusters. { For these results we rerun the scaling runs, but also measure the power on the GPUs during the runs.} On the Alvis cluster, a report with the average power usage for each GPU is automatically generated after each run through the NVIDIA Datacenter GPU Management interface. In addition to this automatic system-provided report, we poll nvidia-smi during the run to { continually measure the power of the GPUs}. With these two measurements, we { check that the average power usage during the run is consistent between the two measurements. As initialization does not properly represent the power draw during a simulation we use the measurements from nvidia-smi to compute the average power usage for the GPUs during the run after initialization.}

On the internal HPE Cray EX system, the power draw of the node is measured using a similar method to that on the Cray XC systems described by \cite{hart2014user}. On the HPE Cray EX system, separate user-readable counters are provided on each node for the CPU, memory, and each of the four accelerator sockets. By measuring the energy counters and the timestamp at the start and end of the job script, a mean power draw for each component can be computed and averaged across the nodes used. We also poll the GPUs with rocm-smi during the run to get an overview of the power draw during the run. With these measurements, { we verify that the average power draw between the two measurements during the course of the run is consistent for the GPUs. We then use the measurements from rocm-smi to compute the average power} usage during the computation, excluding initialization. We note that the differences in methodology mean that some care should be taken when comparing time-averaged power draws between the two GPU systems. 

For our pilot production simulation, we use Alvis at C3SE to generate over a thousand samples of the field and produce over 10TB of flow field data over the course of only 3 days using between 64 and 128 Nvidia A100 GPUs.

%\begin{table}[]
 %   \centering
 %   \begin{tabular}{lll}\toprule
 %      Setup  & Alvis & AMD  \\\midrule
 %      GPU  & A100 HGX & MI250X\\
 %      CPU & 2 Intel Xeon Gold 6338& AMD EPYC 7A53\\
%       DRAM & 512GB DDR4 & 512GB DDR4
%       MPI & OpenMPI 4.0.5 & \\
%       Interconnect & Infiniband(WHICH) & HPE Slingshot(WHICH)\\
   %    Compiler & GCC 10.2& Alistair? \\
  %     GPU Driver & 510.39.01 & Alistair?\\
  %     CUDA/HIP & CUDA 11.1.1 & Alistair?
  %     Bandwidth & 1550 GB/s & 3277 GB/s\\
 %      Peak FP64 & 9.746 TFLOPS & 47.87 TFLOPS\\
 %      TDP & 250W & 500W  \\
 %      \bottomrule \\
 %   \end{tabular}
 %   \caption{Hardware and Software details for our different experimental platforms. Tensor cores omitted from peak FP64.}
%    \label{tbl:exp_setup}
%\end{table}

\begin{figure*}[t]
    \begin{center}
    \scalebox{1}[-1]{\includegraphics[width=1.0\textwidth]{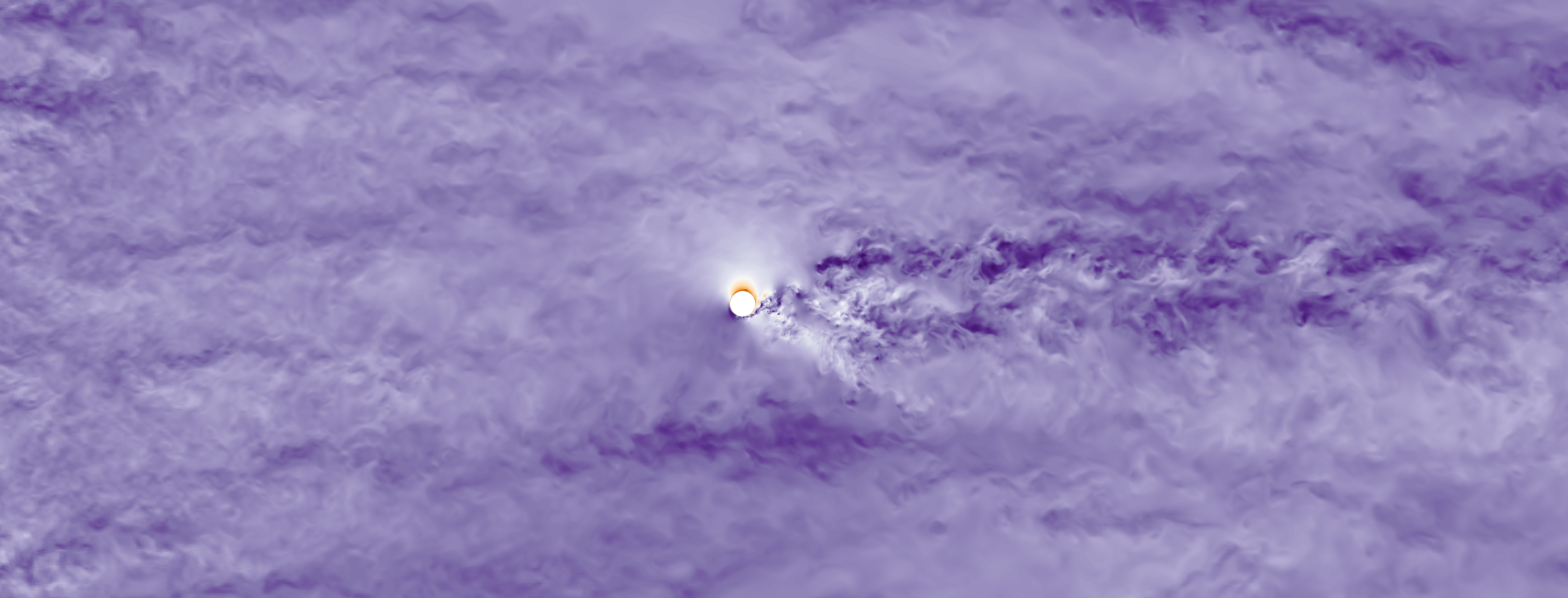}}\\
    \vspace{0.4cm}
    \includegraphics[width=1.0\textwidth]{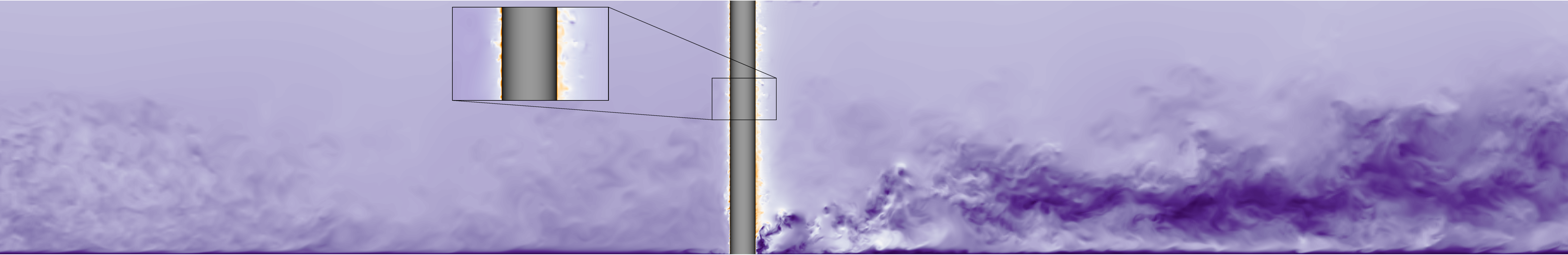}
    \end{center}
    \caption{Two zoomed-in visualizations of the flow around our simulated Flettner rotor. We show the velocity magnitude where a lower velocity is darker violet and a higher velocity becomes white then orange. At the top we show a cut at $z=0$ and at the bottom we show a plane at $y=0.1$. The coordinate axes reflect the mesh as shown in Figure \ref{fig:mesh}, but are zoomed in to better visualize the turbulence close to the cylinder.}
    \label{fig:really_pretty_cylinder}
\end{figure*}

\section{Simulation Results}
After running the simulation for around $15 t^*$ flow time units, where $t^*=t u_{cl}/h$ is the non-dimensional time,  and using around $10~000$ GPU hours, we show the flow field in Figure \ref{fig:really_pretty_cylinder}. We can observe a  significant interaction between the rotor and the turbulent boundary layer, with the wake deflected by the spinning motion. The Flettner rotor generates a spanwise force which could be used to significantly reduce the amount of required thrust of a ship.

A first estimation has been performed by time-averaging the surface integral of the pressure and viscous stress tensor. The resulting aerodynamic force, $\mathbf{F_a}$, is computed as 
\begin{equation}
\mathbf{F_a} = \int_S -p \mathbf{n} + \mathbf{\tau} \mathbf{n} \hspace{1.1mm} \text{d} A 
\end{equation}
where $S$ is the cylinder surface, $p$ the pressure, $\mathbf{n}$ the vector normal to the cylinder surface and $\mathbf{\tau}$ the viscous stress $\tau_{ij}=\mu \partial u_i/\partial u_j$. According to the standard convention, the force is normalized with $0.5 \rho u_{cl} h D$. As a preliminary result, we present the most relevant integral quantities, \emph{i.e.} the drag and lift coefficients. They are the streamwise and spanwise normalized force components respectively. The spanwise lift coefficient $C_l=7.464$ shows an agreement with the experimental data that was measured to be between 7 and 8 for the same $\alpha=3$~(\citealt{bordogna19}). The Reynolds number in our case is smaller, but its influence on $C_l$ seems to be limited. Quite the opposite, the drag coefficient $C_d=1.092$ is strangely dependent on the Reynolds number. Both standard deviations are small (order of magnitude $\approx 10^{-3}$), showing no large temporal fluctuations for these integral estimations. This first pilot investigation shows interesting flow features, as the wake of the cylinder interacts with the boundary layer, encouraging us to continue the study by observing the rotation speed influence or testing different incoming velocity orientations. We aim to highlight new flow characteristics by means of statistical flow analysis and the introduction of modal decomposition, e.g. the proper orthogonal decomposition (POD) to extract the most energetic coherent structures of the flow. More in-depth knowledge of the Flettner rotor dynamics turns out to be extremely relevant in many engineering applications. While our results in this work are preliminary, we have shown how DNS is a valuable tool in understanding how we can leverage Flettner rotors for more sustainable transportation and how DNS can be used to understand flow features not visible in physical experiments.

%\begin{table}[]
%    \centering
%    \begin{tabular}{lll}\toprule
%         Drag coefficient & Lift coefficient\\\midrule
%         $1.092 \pm 0.00038 $ & $7.464 \pm 0.00028$\\ \bottomrule \\
%    \end{tabular}
%    \caption{Our measured lift and drag coefficients and their standard deviation.}
%    \label{tbl:int_quant}
%\end{table}

\section{Performance Results}
For the performance results we compare the two GPU platforms to a CPU baseline by strong scaling with our Flettner rotor case. We also compare the power efficiency of the Nvidia A100 and the AMD  Instinct 250X GPUs.

\begin{figure}
    \centering
    \includegraphics[width=0.48\textwidth]{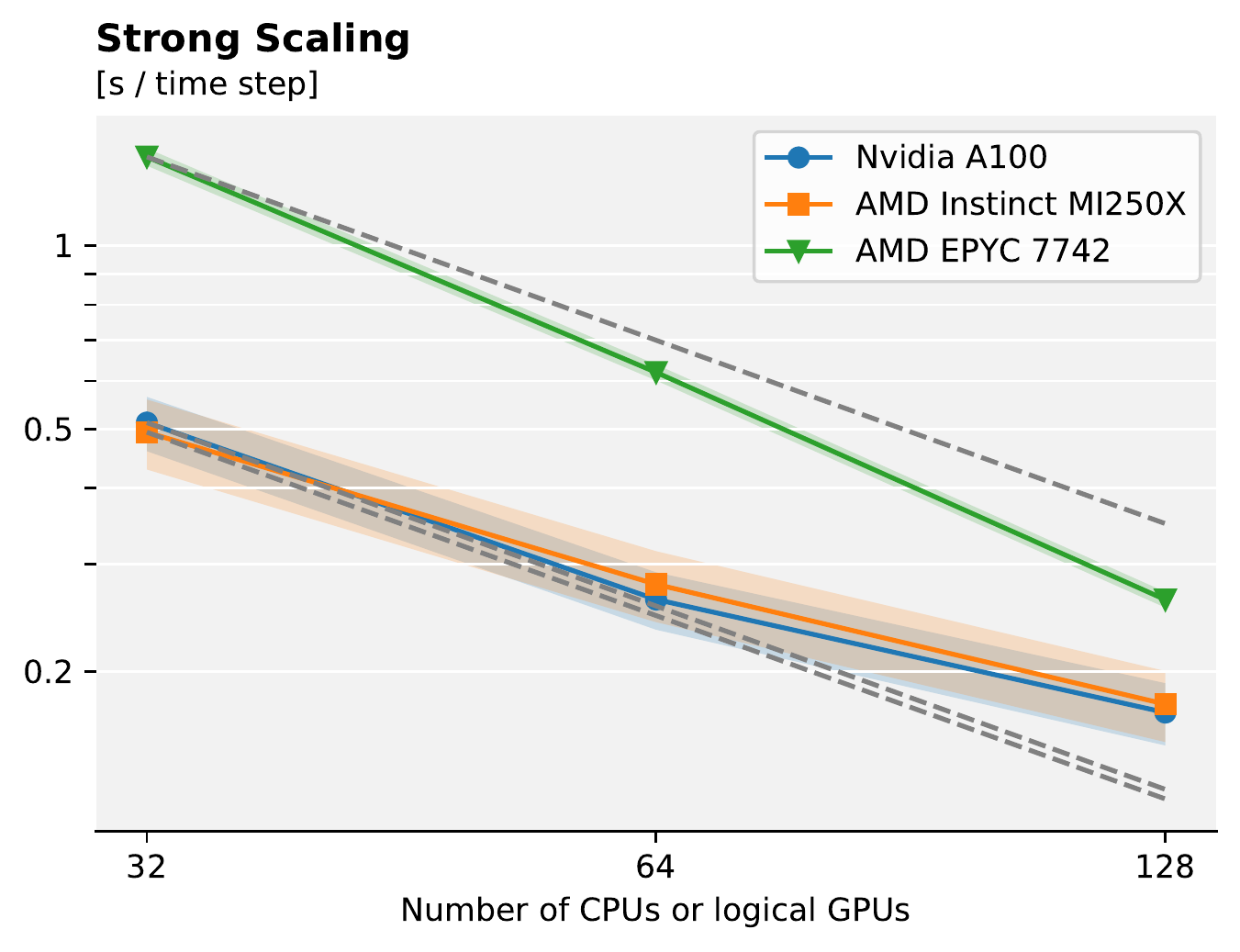}
    \caption{Performance in time per time step as well as shaded areas for the standard deviation. The orange and blue lines represent the GPU systems, while the green line represents the baseline CPU system. We illustrate linear scaling with dotted lines for each platform.}
    \label{fig:scalingresults}
\end{figure}

\begin{figure*}[t]
    \centering
    \includegraphics[width=0.98\textwidth]{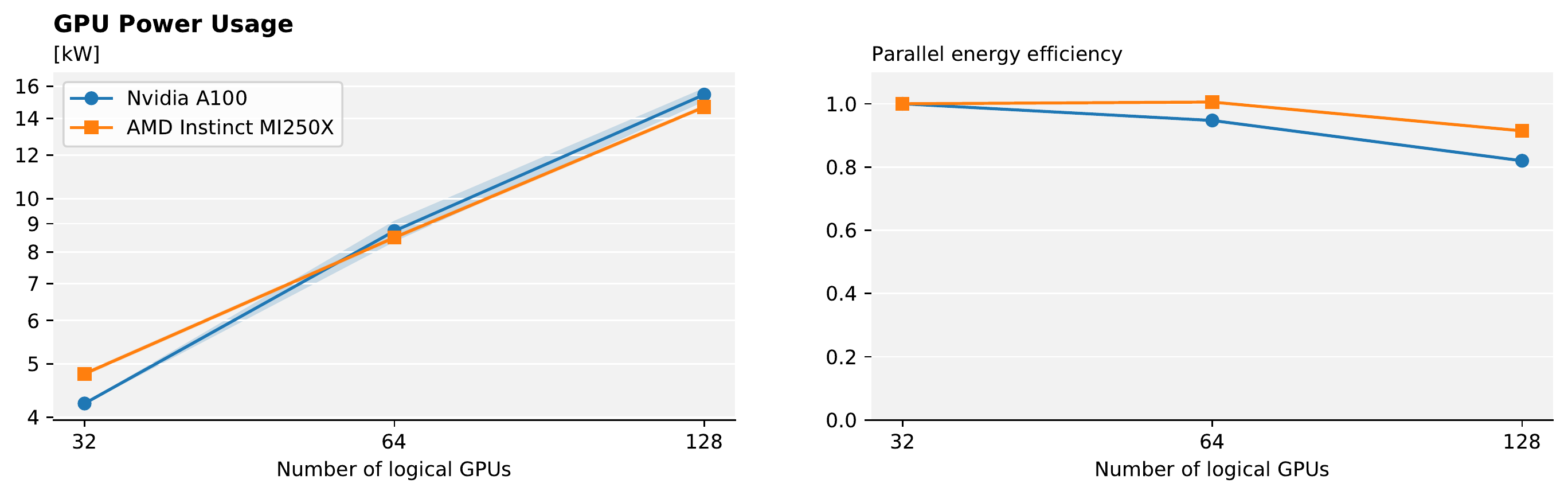}
    \caption{Power usage in watts and the parallel energy efficiency of the GPUs, during a simulation. The shaded area shows the standard deviation between GPUs used in the run for their average power usage. The parallel energy efficiency is computed similarly to the parallel efficiency, but considers energy instead of run time.}
    \label{fig:powerresults}
\end{figure*}
\subsection{Scaling Performance}
In Figure \ref{fig:scalingresults} we present the achieved strong scaling performance for both the AMD Instinct MI250X, Nvidia A100, and the AMD EPYC 7742 CPU baseline. First, comparing the two GPUs, it is clear from the time per time step that { two logical GPUs on the} MI250X correspond to { two} A100s with regards to performance. { We observe that the perfromance of one A100 and one GCD of the MI250X matches very well in all measurements. The average time per time step between the two architectures differs by less than $5\%$, comparing two A100 to one MI250X. This is in contrast to a recent report by \cite{kolev2022ecp} where they observed that one GCD of the MI250X performs closer to $79\%$ of one A100 for a single rod simulation with NekRS, a similar code based on Nek5000~(\citealt{fischer2021nekrs}). They also share our impression during code development that kernels that work well with CUDA/Nvidia do not necessarily map well to HIP/AMD, but need to be modified to properly exploit AMD GPUs}. Comparing the two architectures in Tab. \ref{tbl:exp_setup}, and considering that we expect to be memory-bound as has been discussed in \cite{karp2021High,kolev2021efficient}, we see that the fraction between the bandwidth of two A100 vs one MI250X GPU is 1.05, indicating that they should perform similarly. As many solvers in the wider HPC community are memory-bound, we anticipate that several other codes will obtain similar performance numbers to ours on upcoming GPU supercomputers powered by the AMD Instinct MI250X relative to systems using the Nvidia A100. The relation between the floating-point performance is significantly larger than the observed difference, considering we do not use the tensor cores. The difference in peak FP64 performance suggests that compute-bound applications in FP64 might see larger performance improvements with the MI250X.

We observe a slightly higher standard deviation for the GPUs, compared to the CPU, and this can primarily be attributed to the projection scheme that is used in Neko. This scheme resets every 20-time steps on the GPU, effectively leading to a time variation with a period of 20 for the wall time per time step. On the CPU the projection space is not reset but a more complex scheme where the space is re-orthogonalized is used instead, and we see a lower standard deviation. Other than this difference, the numerical methods used are identical on CPU and GPU.

Comparing the two GPUs to the CPU baseline we see that for a large problem size such as this, the GPUs offer significantly higher performance than modern CPUs at lower node counts. The CPU requires more nodes to achieve the same performance as only a few GPU nodes. In particular, 128 CPUs, totaling 8192 cores performs similarly to only 64 Nvidia A100 GPUs or 32 AMD Instinct MI250X for this specific problem. While the strong scaling of the CPU is higher than the GPUs and running on a large number of cores can offer even higher performance, the possibility of using fewer nodes to get a similar performance is an alluring aspect of the GPUs. For even larger problems with even more elements, these results indicate that GPUs might be the only way to obtain satisfactory performance for smaller computing clusters with a limited number of nodes.

We compute the overall parallel efficiency for the AMD Instinct MI250X and Nvidia A100 to $70\%$ and $75\%$ for { 128 GCDs} and 128 GPUs respectively, with around 3.75 million grid points (7000 elements) per A100 or {{GCD}}. This parallel efficiency is comparable and within one standard deviation. Also, we observe a slightly better average performance for the MI250X with fewer nodes than the A100, adding to the slightly lower parallel efficiency. We anticipate that we will have a satisfactory parallel efficiency until ~3500 elements per logical GPU. For the CPU on the other hand we see a super linear parallel efficiency as the problem size per CPU is very large, between 200 000 grid points (450 elements) per core when using 32 CPUs. There is a significant cache effect as the number of points per core is decreased, leading to its super linear scaling. This super linear behavior for the strong scaling of spectral element codes has been observed previously and is also described by~\cite{fischer2021nekrs,offermans2016strong}. Overall, our parallel efficiency is very high and shows how high-order, matrix-free methods can be used for large-scale DNS, both with conventional CPUs and when GPU acceleration is considered.

It is clear from these scaling results that our problem size is large for the considered number of CPU cores, and that the case requires at least 16k cores to run efficiently on the CPUs, but could use even more, the scaling limit on CPUs has previously been measured to around 50-100 elements per core~(\citealt{offermans2016strong}). Potentially, the CPUs could overtake the GPUs for a large number of cores with regards to performance. However, it is clear that GPU acceleration is necessary to run a case such as this on a fewer number of nodes, yielding a higher performance when the problem size per node is large. For more realistic flow cases, such as simulating the entire rotor ship illustrated in Figure \ref{fig:rotor}, we would have a $Re$ of several million, and thus at least a factor $10^{5}$ increase in number of gridpoints, which for the foreseeable future is out of reach. With upcoming exascale machines we anticipate that GPU acceleration enables simulations at $Re=300~000$, requiring thousands of GPUs. To properly simulate the entire ship configuration, accurate wall-models will be essential to decrease the computational cost as was recently discussed by \cite{bae2022scientific}.

%We strong-scale on both AMD and Nvidia GPUs and show the performance results in Figure \ref{fig:scalingresults}. Comparing the absolute performance between the two different units we see that the absolute performance is comparable between the different units, even though the AMD has a significantly high peak performance in FP64. As the spectral element method has been identified to be primarily limited by data-movement on modern GPUs CITECITECITE and the bandwidth between the GPUs is similar it is tehrefore not surprising that the two different GPUs perform similarly. 

%Regarding the power measurements, we focus on the power usage of the individual GPUs as measured by Nvidia-smi and Craypat. From this we see that the power usage increases with the number of GPUs, but that as the GPU-utilization decreases with a smaller problem per GPU the power consumption of a single GPU also decreases. We can, therefore, when we look at the parallel power efficiency see that it is INSERT PERCENT higher than the parallel perfomance efficiency. We have from previous measurements found that the A100 previously was unmatched with regards to power efficiency for SEM compared to other computer architectures. Now, however,  we see that the MI250 is even more power efficient. While the strong scaling for GPUs is a challenge we see that the actual amount of science per joule produced is currently next to none, even if the strong scaling is sub linear.

\subsection{Energy Efficiency}

We compare the MI250X and A100 GPUs' power consumption in Figure \ref{fig:powerresults}. Here, we see that the MI250X and the A100 perform on par with regard to power efficiency for Neko. The average energy consumed per time step is consistently within $10\%$ between the two GPUs, with a slight edge for the A100 for a smaller number of nodes and the MI250X for a larger number of nodes. { The standard deviation for the average power between GPUs is small (less than 5\%) and the average power measured was highly consistent across GPUs on different nodes during the runs.}  As the AMD MI250X is expected to power several of the largest upcoming supercomputers in the world, the impact of its power efficiency should not be understated, both concerning operating costs and sustainability impact. As the A100 was previously unmatched with regards to power efficiency for this type of application (\citealt{karp2021High}), our results indicate that the AMD MI250X is among the most energy-efficient options for large-scale computational fluid dynamics (CFD) currently available. Further supporting the power efficiency of the A100 and MI250X is that nine of the ten most power efficient supercomputers in the world in November 2021 utilize Nvidia A100 GPUs (\citealt{Green500}). We should note that we omit the energy usage of the network, CPUs, and other peripherals in our power measurements and isolate our focus on only the accelerators. The power measurement methodology also differs and this adds uncertainty to our measurements. The overarching computer system will have a significant impact on the energy efficiency of the computations, but for accelerator heavy systems, the GPUs will draw a significant amount of the overall power.

In Figure \ref{fig:powerresults} we also show the parallel energy efficiency, similar to the usual parallel efficiency, but we consider energy consumed per time step by the GPUs instead of run time. We note that the energy usage per time step that we observe is almost constant as we increase the number of GPUs, unlike the parallel efficiency. For the two GPUs, the parallel energy efficiency stays above $80\%$ with the MI250X being close to or above $90\%$. This indicates, that while GPUs do not offer the same strong scaling characteristics as CPUs, the absolute run time is competitive or better and the energy needed for the GPUs during the simulation is not heavily affected by scaling penalties. It might be more important to consider energy usage per simulation, rather than run time per simulation when comparing different architectures and algorithms. This more correctly corresponds with the operating costs of the computation and sustainability impact. While we have used the power counters available to us through rocm/nvidia-smi, we note that obtaining accurate power usage is a challenge. Currently, direct power meter measurements are not wide-spread and used in production systems and it is an open question how we should compare different  components of heterogenous computer systems at scale with regards to power.

Another aspect of our power measurements is that our simulations are made in double precision. Using lower precision arithmetic to reduce power consumption and increase performance is gaining traction as more and more computing units support lower precision arithmetic. Incorporating other precision formats into Neko is an active area of research, both to increase performance, but also to decrease the energy consumption of large-scale DNS even further. Going forward, Neko will be used to assess the impact of numerical precision on direct numerical simulations. In addition, including adaptive mesh refinement and data-compression techniques to reduce the load on the I/O subsystem will let us use Neko for a wider range of cases and more complex geometries.

%\section{Energy Considerations}
%What do I even want to say here, this feels a bit pseudo-science
%From our performance measurement we see that the total power consumption to execute our simulation takes around X kW/h of electricity. Comparing this to the enery usage of a cargo ship, often above Y kW/h, we have that if our results can reduce the fuel usage by as little as Z\% our simulation has made up for its power consumption much as K times over. While this is only an indication of the impact of our simulation with regards to power efficiency it is of note that these types of simulations cost significant amounts of electricity. A more important aspects than pure run time is therefore the amount of science achieved per unit of energy, something that was also pointed out by Matsouka in CITE. Taking energy into consideration when executing large-scale experiments must therefore also be taken into account.

%Scaling on GPUs 
%GPU power usage, scaling being higher than the observed parallel efficiency
%Some kind of approximation of power needed to rotate cylinder? $\rightarrow$ make napkin math and relate how simulations such as these can help with green transport
%Discussion realting to other computing units?

\section{Related Work}
%Maybe move later? But it might be good to have it here
%Supercomputers power measruement, simulation/science per joule
Our work targets high-fidelity incompressible flow simulations and is based on a spectral element method. The spectral element method was first introduced for CFD by \cite{patera1984spectral,maday1989spectral}. The method has then been developed and derived into several different frameworks such as Nektar++~(\citealt{nektar}) and Nek5000~(\citealt{nek5000}). For DNS, Nek5000 has been well used because of its extreme scalability and accuracy but uses static memory and Fortran 77, making it most suitable for CPUs (\citealt{tufo1999terascale}), although acceleration using OpenACC directives has been considered~(\citealt{otero2019openacc}). Neko shares its roots with Nek5000, but unlike Nek5000, Neko makes use of object-oriented modern Fortran, dynamic memory, and uses a new communication backend. Another approach based on Nek5000, NekRS~(\citealt{fischer2021nekrs}), also provides GPU support, but is rewritten in C++ and uses OCCA extensively to generate code for different backends~(\citealt{medina2014occa}). As our simulation makes use of the openly available KTH Toolboxes originally developed for Nek5000~(\citealt{framework_github}), we could with Neko's device layer make very limited changes to port the necessary parts to Neko and GPUs.

Flettner rotors have recently been studied by conducting massive experimental campaigns in a large wind tunnel~(\citealt{bordogna19,bordogna20}). Numerical simulations allow to set up of a virtual wind tunnel, where the instantaneous solution of the entire flow field is available, unlike wind tunnel experiments where only some local probes or in-plane particle images velocimetry (PIV) measurements are available. Despite the growing interest in this rotor configuration, the literature is limited to RANS simulations see e.g. \cite{demarco16}. In these numerical studies, only the mean flow dynamics is solved and a turbulence model needs to be introduced to close the problem. The presence of a model itself compromises the level of fidelity and accuracy. We present here the first DNS for the flow around a spinning cylinder immersed in a turbulent boundary layer, no periodic boundary conditions are imposed. We do not rely on any model and all the scales are spatially and temporally resolved. While the lift generation mechanism has been well known since the last century (\citealt{magnus1853}), several physical aspects remain unclear, e.g. the inﬂuence of the Reynolds number and wind orientation on the aerodynamic force of the rotating cylinder, and are only possible to study through high-fidelity simulations. 

\section{Conclusion}
We have extended and optimized Neko, a Fortran-based solver with support for modern accelerators, ready for AMD and Nvidia GPUs. We leveraged modern Fortran to use previous tools developed for Fortran 77 and performed the world's first direct numerical simulation of a Flettner rotor in a turbulent boundary layer. We have presented some initial integral quantities and shown that DNS can further our understanding of the turbulent coherent structures and how Flettner rotors can be used to increase sustainability in shipping. We have also presented one of the first performance and energy comparisons between the new AMD Instinct MI250X and Nvidia A100. Overall, we see that modern GPUs enable us to perform relevant, large-scale DNS on relatively few nodes within a time frame and energy budget that was previously not possible.

\begin{acks}
We are grateful to HPE for providing access to their system for this research. We thank C3SE for the opportunity to scale and perform the majority of our simulation using their Alvis system at Chalmers University of Technology in Gothenburg.
\end{acks}

\begin{dci}
The Author(s) declare(s) that there is no conflict of interest.
\end{dci}

\begin{funding}
The author(s) disclosed receipt of the following financial support for the research, authorship, and/or publication of this article: Financial support was provided by the Swedish e-Science Research Centre Exascale Simulation Software Initiative (SESSI); the Swedish Research Council project grant “Efficient Algorithms for Exascale Computational Fluid Dynamics” [grant reference 2019-04723]. Parts of the computations were enabled by resources provided by the Swedish National Infrastructure for Computing (SNIC), partially funded by the Swedish Research Council through grant agreement no. 2018-05973, at C3SE and PDC Centre for High Performance Computing.
\end{funding}

\bibliographystyle{sageh}
\bibliography{references}

\begin{thebibliography}{43}
\providecommand{\natexlab}[1]{#1}
\providecommand{\url}[1]{\texttt{#1}}
\providecommand{\urlprefix}{URL }
\expandafter\ifx\csname urlstyle\endcsname\relax
  \providecommand{\doi}[1]{DOI:\discretionary{}{}{}#1}\else
  \providecommand{\doi}{DOI:\discretionary{}{}{}\begingroup
  \urlstyle{rm}\Url}\fi

\bibitem[{Atzori et~al.(2021)Atzori, Vinuesa, Stroh, Gatti, Frohnapfel and
  Schlatter}]{atzorietal21}
Atzori M, Vinuesa R, Stroh A, Gatti D, Frohnapfel B and Schlatter P (2021)
  Uniform blowing and suction applied to nonuniform adverse-pressure-gradient
  wing boundary layers.
\newblock \emph{Phys. Rev. Fluids} .

\bibitem[{Bae and Koumoutsakos(2022)}]{bae2022scientific}
Bae HJ and Koumoutsakos P (2022) Scientific multi-agent reinforcement learning
  for wall-models of turbulent flows.
\newblock \emph{Nature Communications} 13(1): 1--9.

\bibitem[{Bordogna et~al.(2019)Bordogna, Muggiasca, Giappino, Belloli, Keuning
  and Huijsmans}]{bordogna19}
Bordogna G, Muggiasca S, Giappino S, Belloli M, Keuning J and Huijsmans R
  (2019) Experiments on a {Flettner} rotor at critical and supercritical
  {Reynolds} numbers.
\newblock \emph{Journal of Wind Engineering \& Industrial Aerodynamics} 188:
  193--204.

\bibitem[{Bordogna et~al.(2020)Bordogna, Muggiasca, Giappino, Belloli, Keuning
  and Huijsmans}]{bordogna20}
Bordogna G, Muggiasca S, Giappino S, Belloli M, Keuning J and Huijsmans R
  (2020) The effects of the aerodynamic interaction on the performance of two
  flettner rotors.
\newblock \emph{Journal of Wind Engineering \& Industrial Aerodynamics} 196.

\bibitem[{Cantwell et~al.(2015)Cantwell, Moxey, Comerford, Bolis, Rocco,
  Mengaldo, {De Grazia}, Yakovlev, Lombard, Ekelschot, Jordi, Xu, Mohamied,
  Eskilsson, Nelson, Vos, Biotto, Kirby and Sherwin}]{nektar}
Cantwell C, Moxey D, Comerford A, Bolis A, Rocco G, Mengaldo G, {De Grazia} D,
  Yakovlev S, Lombard JE, Ekelschot D, Jordi B, Xu H, Mohamied Y, Eskilsson C,
  Nelson B, Vos P, Biotto C, Kirby R and Sherwin S (2015) Nektar++: An
  open-source spectral/hp element framework.
\newblock \emph{Computer Physics Communications} 192: 205--219.

\bibitem[{Chevalier et~al.(2007)Chevalier, Lundbladh and
  Henningson}]{Chevalier07}
Chevalier M, Lundbladh A and Henningson D (2007) {SIMSON}–a pseudo-spectral
  solver for incompressible boundary layer flow.
\newblock Technical report.

\bibitem[{De~Marco et~al.(2016)De~Marco, Mancini, Pensa, Calise and
  De~Luca}]{demarco16}
De~Marco A, Mancini S, Pensa C, Calise G and De~Luca F (2016) Flettner rotor
  concept for marine applications: A systematic study.
\newblock \emph{International Journal of Rotating Machinery} .

\bibitem[{Deville et~al.(2002)Deville, Fischer and Mund}]{deville2002high}
Deville MO, Fischer PF and Mund E (2002) \emph{High-order methods for
  incompressible fluid flow}, volume~9.
\newblock Cambridge university press.

\bibitem[{Fischer et~al.(2021)Fischer, Kerkemeier, Min, Lan, Phillips,
  Rathnayake, Merzari, Tomboulides, Karakus, Chalmers
  et~al.}]{fischer2021nekrs}
Fischer P, Kerkemeier S, Min M, Lan YH, Phillips M, Rathnayake T, Merzari E,
  Tomboulides A, Karakus A, Chalmers N et~al. (2021) {NekRS}, a
  {GPU}-accelerated spectral element {Navier-Stokes} solver.
\newblock \emph{arXiv preprint arXiv:2104.05829} .

\bibitem[{Fischer(1998)}]{fischer1998projection}
Fischer PF (1998) Projection techniques for iterative solution of {Ax= b} with
  successive right-hand sides.
\newblock \emph{Computer methods in applied mechanics and engineering}
  163(1-4): 193--204.

\bibitem[{Fischer et~al.(2008)Fischer, Lottes and Kerkemeier}]{nek5000}
Fischer PF, Lottes JW and Kerkemeier SG (2008) {nek5000} {W}eb page.
\newblock \url{http://nek5000.mcs.anl.gov}.

\bibitem[{{Green500}(2021)}]{Green500}
{Green500} (2021) {Green500} {W}eb page.
\newblock \url{https://www.top500.org/lists/green500/}.

\bibitem[{Hart et~al.(2014)Hart, Richardson, Doleschal, Ilsche, Bielert and
  Kappel}]{hart2014user}
Hart A, Richardson H, Doleschal J, Ilsche T, Bielert M and Kappel M (2014)
  User-level power monitoring and application performance on cray xc30
  supercomputers.
\newblock \emph{Proceedings of the Cray User Group (CUG)} 1.

\bibitem[{Jansson(2021)}]{hpcasia21}
Jansson N (2021) {Spectral Element Simulations on the NEC SX-Aurora TSUBASA}.
\newblock In: \emph{{The International Conference on High Performance Computing
  in Asia-Pacific Region}}, HPC Asia 2021. New York, NY, USA: ACM, pp. 32--39.

\bibitem[{Jansson et~al.(2021)Jansson, Karp, Podobas, Markidis and
  Schlatter}]{jansson2021neko}
Jansson N, Karp M, Podobas A, Markidis S and Schlatter P (2021) Neko: A modern,
  portable, and scalable framework for high-fidelity computational fluid
  dynamics.
\newblock \emph{arXiv preprint arXiv:2107.01243} .

\bibitem[{Karniadakis et~al.(1991)Karniadakis, Israeli and
  Orszag}]{karniadakis1991high}
Karniadakis GE, Israeli M and Orszag SA (1991) High-order splitting methods for
  the incompressible {Navier-Stokes} equations.
\newblock \emph{Journal of computational physics} 97(2): 414--443.

\bibitem[{Karp et~al.(2021)Karp, Podobas, Jansson, Kenter, Plessl, Schlatter
  and Markidis}]{karp2021High}
Karp M, Podobas A, Jansson N, Kenter T, Plessl C, Schlatter P and Markidis S
  (2021) High-performance spectral element methods on field-programmable gate
  arrays : Implementation, evaluation, and future projection.
\newblock In: \emph{2021 IEEE International Parallel and Distributed Processing
  Symposium (IPDPS)}. pp. 1077--1086.
\newblock \doi{10.1109/IPDPS49936.2021.00116}.

\bibitem[{Karp et~al.(2022{\natexlab{a}})Karp, Podobas, Jansson, Schlatter and
  Markidis}]{karp2022reducing}
Karp M, Podobas A, Jansson N, Schlatter P and Markidis S (2022{\natexlab{a}})
  Reducing communication in the conjugate gradient method: A case study on
  high-order finite elements.
\newblock In: \emph{Platform for Advanced Scientific Computing Conference (PASC
  '22)}.
\newblock \doi{10.1145/3539781.3539785}.

\bibitem[{Karp et~al.(2022{\natexlab{b}})Karp, Podobas, Kenter, Jansson,
  Plessl, Schlatter and Markidis}]{karp2022HpcAsia}
Karp M, Podobas A, Kenter T, Jansson N, Plessl C, Schlatter P and Markidis S
  (2022{\natexlab{b}}) A high-fidelity flow solver for unstructured meshes on
  field-programmable gate arrays: Design, evaluation, and future challenges.
\newblock In: \emph{International Conference on High Performance Computing in
  Asia-Pacific Region}, HPCAsia2022. New York, NY, USA: Association for
  Computing Machinery.
\newblock ISBN 9781450384988, p. 125–136.
\newblock \doi{10.1145/3492805.3492808}.

\bibitem[{Keryell et~al.(2015)Keryell, Reyes and Howes}]{keryell2015khronos}
Keryell R, Reyes R and Howes L (2015) Khronos sycl for opencl: a tutorial.
\newblock In: \emph{Proceedings of the 3rd International Workshop on OpenCL}.
  pp. 1--1.

\bibitem[{Kolev et~al.(2022)Kolev, Fischer, Abdelfattah, Beams, Brown, Camier,
  Carson, Chalmers, Dobrev, Dudouit et~al.}]{kolev2022ecp}
Kolev T, Fischer P, Abdelfattah A, Beams N, Brown J, Camier JS, Carson R,
  Chalmers N, Dobrev V, Dudouit Y et~al. (2022) Ecp milestone report high-order
  algorithmic developments and optimizations for more robust exascale
  applications wbs 2.2. 6.06, milestone ceed-ms38 .

\bibitem[{Kolev et~al.(2021)Kolev, Fischer, Min, Dongarra, Brown, Dobrev,
  Warburton, Tomov, Shephard, Abdelfattah et~al.}]{kolev2021efficient}
Kolev T, Fischer P, Min M, Dongarra J, Brown J, Dobrev V, Warburton T, Tomov S,
  Shephard MS, Abdelfattah A et~al. (2021) Efficient exascale discretizations:
  High-order finite element methods.
\newblock \emph{The International Journal of High Performance Computing
  Applications} 35(6): 527--552.

\bibitem[{Lottes and Fischer(2005)}]{lottes2005hybrid}
Lottes JW and Fischer PF (2005) Hybrid multigrid/schwarz algorithms for the
  spectral element method.
\newblock \emph{Journal of Scientific Computing} 24(1): 45--78.

\bibitem[{Maday and Patera(1989)}]{maday1989spectral}
Maday Y and Patera AT (1989) Spectral element methods for the incompressible{
  Navier-Stokes} equations.
\newblock \emph{State-of-the-art surveys on computational mechanics (A90-47176
  21-64). New York} : 71--143.

\bibitem[{Magnus(1853)}]{magnus1853}
Magnus G (1853) {Über die Abweichung der Geschosse, und: Über eine abfallende
  Erscheinung bei rotierenden Körpern}.
\newblock \emph{Annalen der Physik} 1: 1--29.

\bibitem[{Medina et~al.(2014)Medina, St-Cyr and Warburton}]{medina2014occa}
Medina DS, St-Cyr A and Warburton T (2014) Occa: A unified approach to
  multi-threading languages.
\newblock \emph{arXiv preprint arXiv:1403.0968} .

\bibitem[{Offermans et~al.(2016)Offermans, Marin, Schanen, Gong, Fischer,
  Schlatter, Obabko, Peplinski, Hutchinson and Merzari}]{offermans2016strong}
Offermans N, Marin O, Schanen M, Gong J, Fischer P, Schlatter P, Obabko A,
  Peplinski A, Hutchinson M and Merzari E (2016) On the strong scaling of the
  spectral element solver nek5000 on petascale systems.
\newblock In: \emph{Proceedings of the Exascale Applications and Software
  Conference 2016}. pp. 1--10.

\bibitem[{Orszag(1979)}]{orszag1979spectral}
Orszag SA (1979) Spectral methods for problems in complex geometrics.
\newblock In: \emph{Numerical methods for partial differential equations}.
  Elsevier, pp. 273--305.

\bibitem[{Orszag et~al.(1986)Orszag, Israeli and Deville}]{orszag1986boundary}
Orszag SA, Israeli M and Deville MO (1986) Boundary conditions for
  incompressible flows.
\newblock \emph{Journal of Scientific Computing} 1(1): 75--111.

\bibitem[{Otero et~al.(2019)Otero, Gong, Min, Fischer, Schlatter and
  Laure}]{otero2019openacc}
Otero E, Gong J, Min M, Fischer P, Schlatter P and Laure E (2019) Openacc
  acceleration for the {PN--PN-2} algorithm in nek5000.
\newblock \emph{Journal of Parallel and Distributed Computing} 132: 69--78.

\bibitem[{Pachauri et~al.(2007)Pachauri, Reisinger et~al.}]{pachauri2007ipcc}
Pachauri RK, Reisinger A et~al. (2007) {IPCC} fourth assessment report.
\newblock \emph{IPCC, Geneva} 2007.

\bibitem[{Patera(1984)}]{patera1984spectral}
Patera AT (1984) A spectral element method for fluid dynamics: laminar flow in
  a channel expansion.
\newblock \emph{Journal of Computational Physics} 54(3): 468--488.

\bibitem[{Peplinski(2022)}]{framework_github}
Peplinski A (2022) {KTH Framework for Nek5000}.
\newblock \url{https://github.com/KTH-Nek5000/KTH\textunderscore Framework}.

\bibitem[{Reid(2008)}]{reid2008new}
Reid J (2008) The new features of {Fortran} 2008.
\newblock In: \emph{ACM SIGPLAN Fortran Forum}, volume~27. ACM New York, NY,
  USA, pp. 8--21.

\bibitem[{Saad and Schultz(1986)}]{saad1986gmres}
Saad Y and Schultz MH (1986) {GMRES}: A generalized minimal residual algorithm
  for solving nonsymmetric linear systems.
\newblock \emph{SIAM Journal on scientific and statistical computing} 7(3):
  856--869.

\bibitem[{Schlatter and Örlü(2012)}]{schlatter12}
Schlatter P and Örlü R (2012) Turbulent boundary layers at moderate reynolds
  numbers: Inflow length and tripping effects.
\newblock \emph{J. of Fluid Mech.} 710: 5--34.

\bibitem[{Seddiek and Ammar(2021)}]{seddiek2021harnessing}
Seddiek IS and Ammar NR (2021) {Harnessing wind energy on merchant ships: case
  study Flettner rotors onboard bulk carriers}.
\newblock \emph{Environmental Science and Pollution Research} 28(25):
  32695--32707.

\bibitem[{Seifert(2012)}]{seifert2012review}
Seifert J (2012) A review of the {Magnus} effect in aeronautics.
\newblock \emph{Progress in Aerospace Sciences} 55: 17--45.

\bibitem[{Smith et~al.(2015)Smith, Traut, Bows-Larkin, Anderson, McGlade and
  Wrobel}]{smith2015co2}
Smith T, Traut M, Bows-Larkin A, Anderson K, McGlade C and Wrobel P (2015)
  {CO2} targets, trajectories and trends for international shipping .

\bibitem[{{\'S}wirydowicz et~al.(2019){\'S}wirydowicz, Chalmers, Karakus and
  Warburton}]{swirydowicz2019acceleration}
{\'S}wirydowicz K, Chalmers N, Karakus A and Warburton T (2019) Acceleration of
  tensor-product operations for high-order finite element methods.
\newblock \emph{The International Journal of High Performance Computing
  Applications} 33(4): 735--757.

\bibitem[{Tanarro et~al.(2019)Tanarro, Vinuesa and Schlatter}]{tanarroetal19}
Tanarro A, Vinuesa R and Schlatter P (2019) Effect of adverse pressure
  gradients on turbulent wing boundary layers.
\newblock \emph{J. of Fluid Mech.} 883.

\bibitem[{Tufo and Fischer(1999)}]{tufo1999terascale}
Tufo HM and Fischer PF (1999) Terascale spectral element algorithms and
  implementations.
\newblock In: \emph{Proceedings of the 1999 ACM/IEEE Conference on
  Supercomputing}. pp. 68--es.

\bibitem[{Wienke et~al.(2012)Wienke, Springer, Terboven
  et~al.}]{wienke2012openacc}
Wienke S, Springer P, Terboven C et~al. (2012) Openacc—first experiences with
  real-world applications.
\newblock In: \emph{European Conference on Parallel Processing}. Springer, pp.
  859--870.

\end{thebibliography}

\end{document}